\documentclass[12pt,a4paper]{article}

\usepackage{a4wide}
\usepackage{amsmath}
\usepackage{bm}
\usepackage{amssymb}
\usepackage{hyperref}
\usepackage{epic}

\newcommand{\Op}{\mathcal{O}}

\newcommand{\cN}{{\cal N}}

\newcommand{\CA}{{C_A}}
\newcommand{\CF}{{C_F}}
\newcommand{\nf}{{n_f}}
\newcommand{\TF}{{T_F}}
\newcommand{\NF}{{N_F}}
\newcommand{\als}{{a_s}}

\def\CAP#1{{{C_A^#1}}}
\def\CFP#1{{{C_F^#1}}}
\def\TFP#1{{{T_F^#1}}}
\def\nfP#1{{{n_f^#1}}}
\def\z#1{{{\zeta_#1}}}
\newcommand{\pslash}{p \! \! \! /}

\begin{document}

\thispagestyle{empty}

\begin{flushright}\footnotesize
\texttt{HU-Mathematik-2014-31}\\
\texttt{HU-EP-14/47}\\
\vspace{0.5cm}
\end{flushright}

\begin{center}
{\Large{\bf
Four loop anomalous dimension \\[1mm]
of the third and fourth~moments \\[3mm]
of the non-singlet twist-2 operator in QCD
}}
\vspace{15mm}

{\sc
V.~N.~Velizhanin}\\[5mm]

{\it Institut f{\"u}r  Mathematik und Institut f{\"u}r Physik\\
Humboldt-Universit{\"a}t zu Berlin\\
IRIS Adlershof, Zum Gro\ss{}en Windkanal 6\\
12489 Berlin, Germany\\
and
}\\
{\it Theoretical Physics Division\\
Petersburg Nuclear Physics Institute\\
Orlova Roscha, Gatchina\\
188300 St.~Petersburg, Russia}\\[5mm]

\textbf{Abstract}\\[2mm]
\end{center}

\noindent{
We present the result of a calculation for the third and fourth moments of the non-singlet four-loop
anomalous dimension of Wilson twist-2 operators in QCD with full color and flavour structures.
We discuss also a general expressions for some contributions to the full four-loop anomalous dimension obtained with the help of the method, based on LLL-algorithm,
which was proposed earlier by us for the reconstruction of a general form of the anomalous dimension from a fixed values.
}
\newpage

\setcounter{page}{1}

In our recent paper~\cite{Velizhanin:2014zla} we calculated the higher moments for the non-planar contribution to the four-loop anomalous dimension of the twist-2 operators in the maximally extended ${\mathcal N}=4$ Supersymmetric Yang-Mills (SYM) theory. As ${\mathcal N}=4$ SYM theory is a generalization of the Quantum Chromodynamics (QCD) we have all the necessary to perform a similar calculations for the four-loop anomalous dimension in QCD. In this paper we present the results for the third and fourth moments of the non-singlet four-loop anomalous dimension of Wilson twist-2 operators in QCD with the full color and flavour structures, which extend our previous result~\cite{Velizhanin:2011es} to the next two moments.
All general formal equations can be found, for example, in Ref.~\cite{Larin:1993vu}. So, here we will only give the details, which are concerned the calculations of the third and fourth moments together with the discussion of a general results.

The standard twist-2 irreducible (i.e. symmetrical and traceless in indices
$\mu_1\ldots\mu_N$, what usually denoting with curly brackets) flavour non-singlet quark operators with the Lorenz spin-$N$ have the following form:
\begin{equation}
\Op^{a,\{\mu_1\ldots\mu_N\}}=\bar\psi\lambda^a
\gamma^{\{\mu_1}{\mathcal D}^{\mu_2}\ldots {\mathcal D}^{\mu_N\}}\psi,\qquad a=1,2,\ldots,(n_f^2-1),
\label{NSOpN}
\end{equation}
where ${\mathcal D}^{\mu_j}$ are the covariant derivatives, $\lambda^a$ are the generators of the flavour group $SU(n_f)$.
In general there are three independent quark non-singlet distributions ${\rm{q}}_{{\rm ns},ij}^{\pm}$ and ${\rm{q}}_{{\rm ns}}^{\rm{V}}$ and one quark singlet distribution ${\rm{q}}_{\rm{s}}$.
The three non-singlet distributions are the flavour asymmetries
\begin{eqnarray}
\label{ns-dens}
{\rm{q}}_{{\rm ns},ij}^{\pm} = {\rm{q}}_i \pm \Bar{{\rm{q}}}_i - ({\rm{q}}_j \pm \Bar{{\rm{q}}}_j)\, ,
\end{eqnarray}
and the sum of the valence distributions of all flavours,
\begin{eqnarray}
\label{v-dens}
{\rm{q}}_{\rm ns}^{\rm{V}} = \sum\limits_{i=1}^{n_f} ({\rm{q}}_i - \Bar{{\rm{q}}}_i) \, ,
\end{eqnarray}
while the singlet distribution is simply the sum of the distributions of all flavours,
\begin{eqnarray}
\label{s-dens}
{\rm{q}}_{\rm s} = \sum\limits_{i=1}^{n_f} ({\rm{q}}_i + \Bar{{\rm{q}}}_i) \, .
\end{eqnarray}
Here we will calculate the anomalous dimensions $\gamma^{\pm}_{\mathrm {NS}}$ related only with ${\rm{q}}_{{\mathrm{NS}},ij}^{\pm}$ and upper/lower sign corresponds to the operators
$\Op^{a,\{\mu_1\ldots\mu_N\}}$ with even/odd $N$.

So, in this paper we will interesting to the operators with $N=3$ (third moment) and $N=4$ (fourth moment). The results up to three-loop can be found in Refs.~\cite{Gross:1973ju,Floratos:1977au,GonzalezArroyo:1979df,Larin:1993vu,Retey:2000nq}. The Feynman rules for the operator $\mathcal{O}_{\mu _{1},...,\mu _{N}}^{\lambda }$ with the different number of the gauge fields can be found in Ref.~\cite{Bierenbaum:2009mv} and the projectors can be found in Ref.~\cite{Bierenbaum:2009mv} (see also~\cite{Gracey:2006zr}). Note, that for the operator with $N=4$ the operator vertex with the five lines is appeared (two quark and three gluons lines).

As in our previous papers the calculations of diagrams were performed with \texttt{FORM}~\cite{Vermaseren:2000nd}, using \texttt{FORM} package \texttt{COLOR}~\cite{vanRitbergen:1998pn} for evaluation of the color traces. For the dealing with a huge number of diagrams we use a program \texttt{DIANA}~\cite{Tentyukov:1999is}, which call \texttt{QGRAF}~\cite{Nogueira:1991ex} to generate all diagrams.
The reduction of all appeared scalar integrals to the master-integrals was done with our program \texttt{BAMBA} based on the algorithm of Laporta~\cite{Laporta:2001dd}.
We used, following Ref.~\cite{Vladimirov:1979zm} (see also Refs.~\cite{Misiak:1994zw,Chetyrkin:1997fm,Czakon:2004bu}  for details), the infra-red rearrangement (IRR) procedure which reduce a propagator-type diagrams to a fully massive tadpole diagrams.
For the reduction of the newly appeared Feynman integrals with a higher powers of denominators and numerators we considerably improved our {\texttt{MATHEMATICA}} code \texttt{BAMBA}.
Practically, all computations are divided into three parts: calculation of the planar-based diagrams, calculation of the non-planar-based diagrams and a renormalization.

For the calculations we used the following forms of the non-singlet operators (\ref{NSOpN}) for the third ($N=3$) and fourth moment ($N=4$)
\begin{eqnarray}
{\mathcal O}^{a,\mu\nu\rho}_{\mathrm {NS}}& =& \bar{\psi} \lambda^a\gamma^{\mu} {\mathcal D}^\nu{\mathcal D}^{\rho}\psi\,,\label{NSOp3}\\
{\mathcal O}^{a,\mu\nu\rho\sigma}_{\mathrm {NS}}& =& \bar{\psi} \lambda^a\gamma^{\mu} {\mathcal D}^\nu{\mathcal D}^{\rho}{\mathcal D}^{\sigma}\psi\label{NSOp4}
\end{eqnarray}
and we don't need a symmetrization and a subtraction of the traces as we will multiply operators to the corresponding projectors, which are the symmetric traceless tensors .

The projectors for these operators can be found in Ref.~\cite{Bierenbaum:2009mv} and explicitly they have the following forms:
\begin{eqnarray}
\Pi_{\mu\nu\rho} &=& \frac{1}{D-1} \left[
\frac{D+2}{p^6}\,p_{\mu}p_{\nu}p_{\rho}
-\frac{1}{p^4}\,g_{\{\mu\nu}p_{\rho\}}
\right]\frac{\pslash}{4}\, ,\label{ProjectorN3}\\
\Pi_{\mu\nu\rho\sigma} &=&
\frac{1}{D^2-1}\left[
\frac{(D+4)(D+2)}{p^8}\,p_{\mu}p_{\nu}p_{\rho}p_{\sigma}
-\frac{D+2}{p^6}\,g_{\{\mu\nu}p_{\rho}p_{\sigma\}}
+\frac{1}{p^4}\,g_{\{\mu\nu}g_{\rho\sigma\}}
\right]\frac{\pslash}{4}\,.\label{ProjectorN4}
\end{eqnarray}

For the renormalization we need the three-loop renormalization constants for the operator insertion $g\bar{\psi}\lambda^a \gamma^{\{\mu} {\mathcal A}^{\nu}{}^{\, \ldots\}} \psi$ with two quarks and one gluon legs, $g\bar{\psi}\lambda^a \gamma^{\{\mu} {\mathcal A}^{\nu}{\mathcal A}^{\rho}{}^{\, \ldots\}} \psi$ with two quarks and two gluon legs and $g\bar{\psi}\lambda^a \gamma^{\{\mu} {\mathcal A}^{\nu}{\mathcal A}^{\rho} {\mathcal A}^{\sigma\}} \psi$ with two quarks and three gluon legs, which can be obtained order by order in a usual way from the renormalization of the corresponding operator ${\mathcal O}^{a,\{\mu\nu\, \ldots\}}_{\mathrm {NS}}$ with the same number of Lorentz indices as
\begin{eqnarray}
Z_{\bar{\psi}\lambda^a \gamma^{\{\mu} {\mathcal A}^{\nu}{}^{\,\ldots\}} \psi}&=&Z_{{\mathcal O}^{a,\{\mu\nu\, \ldots\}}_{\mathrm {NS}}}Z_{{\mathcal A}}^{1/2}Z_g^{1/2}Z_\psi\,,\\
Z_{\bar{\psi}\lambda^a \gamma^{\{\mu} {\mathcal A}^{\nu}{\mathcal A}^{\rho}{}^{\,\ldots\}} \psi}&=&Z_{{\mathcal O}^{a,\{\mu\nu\rho\, \ldots\}}_{\mathrm {NS}}}Z_{{\mathcal A}}Z_gZ_\psi\,,\\
Z_{\bar{\psi}\lambda^a \gamma^{\{\mu} {\mathcal A}^{\nu}{\mathcal A}^{\rho}{\mathcal A}^{\sigma}{}^{\,\ldots\}} \psi}&=&Z_{{\mathcal O}^{a,\{\mu\nu\rho\sigma\, \ldots\}}_{\mathrm {NS}}}Z_{{\mathcal A}}^{3/2}Z_g^{3/2}Z_\psi\,,
\end{eqnarray}
where $Z_{{\mathcal A}}$, $Z_g$ and $Z_\psi$ are the renormalization constants for gluon filed ${\mathcal A}^\mu$, coupling constant and quark correspondingly.

Our final result is \footnote{The result for $N\!=\!2$~\cite{Velizhanin:2011es} contains a misprint in the last term of Eq.~(22): $128$ should be replaced by $64$.} \footnote{The results for $N=2,\, 3,\, 4$ can be found as the ancillary file of the arXiv version this paper.}:
\begin{eqnarray}
\gamma^{4-loop}_{{\mathrm{NS}}}(3)&=&
\als\frac{25}{6} {\CF}
+\als^2
\bigg[
\frac{535}{27} {\CF} {\CA }
-\frac{2035}{432} {\CFP2}
-\frac{415}{54} {\CF} {\TF \nf}
\bigg]\nonumber\\&&\hspace*{-16mm}
+\,\als^3
\bigg[
{\CFP3}
\left(
\frac{110 {\z3}}{3}
-\frac{244505}{15552}
\right)
+ {\CFP2}{\CA }
\left(
-55 {\z3}
-\frac{311213}{15552}
\right)\nonumber\\&&\hspace*{-3mm}
+{\CFP2} {\TF \nf}
\left(
\frac{200}{3} {\z3}
-\frac{203627}{3888}
\right)
+{\CF}\CAP2
\left(
\frac{55}{3} {\z3}
+\frac{889433}{7776}
\right)\nonumber\\&&\hspace*{-3mm}
+ {\CF}{\CA } {\TF \nf}
\left(
-\frac{200}{3} {\z3}
-\frac{62249}{1944}
\right)
-\frac{2569}{486} {\CF} {\TFP2 \nfP2}
\bigg]\nonumber\\&&\hspace*{-16mm}
+\als^4 \bigg[
{\CFP4}
\left(
\frac{24380 }{81}{\z3}
-\frac{2000}{3} {\z5}
+\frac{341611945}{2239488}
\right)\nonumber\\&&\hspace*{-3mm}
+{\CFP3} {\TF \nf}
\left(
\frac{6986}{81}{\z3}
+\frac{220}{3}{\z4}
-\frac{2000}{3}{\z5}
+\frac{38386673}{139968}
\right)\nonumber\\&&\hspace*{-3mm}
+ {\CFP3}{\CA }
\left(
-\frac{140057}{648}{\z3}
-\frac{605}{3}{\z4}
+\frac{2900}{3}{\z5}
+\frac{40709323}{279936}
\right)\nonumber\\&&\hspace*{-3mm}
+{\CFP2} {\TFP2 \nfP2}
\left(
-\frac{2440}{9}{\z3}
+\frac{400}{3}{\z4}
+\frac{1313443}{17496}
\right)\nonumber\\&&\hspace*{-3mm}
+ {\CFP2}{\CA } {\TF \nf}
\left(
\frac{31547}{27}{\z3}
-\frac{1430}{3}{\z4}
+\frac{1000}{9}{\z5}
-\frac{22941613}{69984}
\right)\nonumber\\&&\hspace*{-3mm}
+ {\CFP2}{\CAP2}
\left(
-\frac{4843}{27}{\z3}
+\frac{605}{2}{\z4}
-\frac{1325}{9}{\z5}
-\frac{503877829}{559872}
\right)\nonumber\\&&\hspace*{-3mm}
+ {\CF}{\CAP3}
\left(
\frac{125219}{648}{\z3}
-\frac{605}{6}{\z4}
-\frac{5950}{27}{\z5}
+\frac{72667541}{69984}
\right)\nonumber\\&&\hspace*{-3mm}
+{\CF} {\CAP2} {\TF \nf}
\left(
-\frac{11483}{9}{\z3}
+\frac{1210}{3}{\z4}
+\frac{14000}{27}{\z5}
-\frac{2366971}{3888}
\right)\nonumber\\&&\hspace*{-3mm}
+ {\CF} {\CA }{\TFP2 \nfP2}
\left(
\frac{2440}{9}{\z3}
-\frac{400}{3}{\z4}
+\frac{79747}{1458}
\right)
\nonumber\\&&\hspace*{-3mm}
+{\CF} {\TFP3 \nfP3}
\left(
\frac{1600}{81}{\z3}
-\frac{23587}{4374}
\right)\nonumber\\&&\hspace*{-3mm}
+\frac{5}{9}\frac{ d_F^{abcd}d_A^{abcd}}{\NF}
\left(
1520{\z3}
-1460{\z5}
-51
\right)\nonumber\\&&\hspace*{-3mm}
+\nf\frac{5}{9}\frac{ d_F^{abcd}d_F^{abcd}}{\NF}
\left(
392{\z3}
-800{\z5}
+165
\right)
\bigg],\label{ADM3}
\end{eqnarray}
\begin{eqnarray}
\gamma^{4-loop}_{{\mathrm{NS}}}(4)&=&
\als\frac{157}{30} {\CF}
+\als^2
\bigg[
\frac{16157}{675} {\CA } {\CF}
-\frac{287303}{54000} {\CFP2}
-\frac{13271}{1350} {\CF} {\TF \nf}
\bigg]\nonumber\\&&\hspace*{-16mm}
+\als^3
\bigg[
{\CFP3}
\left(
\frac{2878}{75}{\z3}
-\frac{714245693}{48600000}
\right)
+{\CFP2}{\CA }
\left(
-\frac{1439}{25}{\z3}
-\frac{267028157}{9720000}
\right)\nonumber\\&&\hspace*{-3mm}
+{\CFP2} {\TF \nf}
\left(
\frac{1256}{15}{\z3}
-\frac{165237563}{2430000}
\right)
+{\CF}{\CAP2}
\left(
\frac{1439}{75}{\z3}
+\frac{136066373}{972000}
\right)\nonumber\\&&\hspace*{-3mm}
+ {\CF}{\CA } {\TF \nf}
\left(
-\frac{1256}{15}{\z3}
-\frac{8802581}{243000}
\right)
-\frac{384277 }{60750}{\CF} {\TFP2 \nfP2}
\bigg]\nonumber\\&&\hspace*{-16mm}
+\als^4
\bigg[
{\CFP4}
\left(
\frac{14504764}{50625}{\z3}
-\frac{25136}{45}{\z5}
+\frac{3482407012657}{34992000000}
\right)\nonumber\\&&\hspace*{-3mm}
+{\CFP3} {\TF \nf}
\left(
\frac{1641922}{10125}{\z3}
+\frac{5756}{75}{\z4}
-\frac{2512}{3}{\z5}
+\frac{29581840417}{87480000}
\right)\nonumber\\&&\hspace*{-3mm}
+ {\CFP3}{\CA }
\left(
-\frac{10215349}{81000}{\z3}
-\frac{15829}{75}{\z4}
+\frac{33004}{45}{\z5}
+\frac{33802068299}{174960000}
\right)\nonumber\\&&\hspace*{-3mm}
+ {\CFP2}{\CAP2}
\left(
-\frac{2497339}{16875}{\z3}
+\frac{15829 }{50}{\z4}
-\frac{1645}{9}{\z5}
-\frac{1557367902137}{1749600000}
\right)\nonumber\\&&\hspace*{-3mm}
+{\CFP2}{\CA } {\TF \nf}
\left(
\frac{4588639}{3375}{\z3}
-\frac{43174}{75}{\z4}
+\frac{1256}{9}{\z5}
-\frac{89325051233}{218700000}
\right)\nonumber\\&&\hspace*{-3mm}
+{\CFP2} {\TFP2 \nfP2}
\left(
-\frac{8584}{25}{\z3}
+\frac{2512}{15}{\z4}
+\frac{5419760639}{54675000}
\right)\nonumber\\&&\hspace*{-3mm}
+{\CF}{\CAP3}
\left(
\frac{13461191}{81000}{\z3}
-\frac{15829 }{150}{\z4}
-\frac{18646}{135}{\z5}
+\frac{49455970561}{43740000}
\right)\nonumber\\&&\hspace*{-3mm}
+ {\CF}{\CAP2} {\TF \nf}
\left(
-\frac{5247961}{3375}{\z3}
+\frac{37418}{75}{\z4}
+\frac{87472}{135}{\z5}
-\frac{1796654459}{2430000}
\right)\nonumber\\&&\hspace*{-3mm}
+{\CF}{\CA }  {\TFP2 \nfP2}
\left(
\frac{8584}{25}{\z3}
-\frac{2512}{15}{\z4}
+\frac{60167591}{911250}
\right)\nonumber\\&&\hspace*{-3mm}
+{\CF} {\TFP3 \nfP3}
\left(
\frac{10048}{405}{\z3}
-\frac{17813699}{2733750}
\right)\nonumber\\&&\hspace*{-3mm}
+\frac{ d_F^{abcd}d_A^{abcd}}{\NF}
\left(
\frac{63568}{45}{\z3}
-\frac{78868}{45}{\z5}
+\frac{254713}{1350}
\right)\nonumber\\&&\hspace*{-3mm}
+\nf\frac{ d_F^{abcd}d_F^{abcd}}{\NF}
\left(
\frac{22552}{75}{\z3}
-\frac{26912}{45}{\z5}
+\frac{16568}{135}
\right)
\bigg],\label{ADM4}
\end{eqnarray}
where $\NF$ is the dimension of the fermion representation (i.e. the number of quark colours), $\nf$ is the number of quark flavors and for the color group $SU(N_c)$ (see Ref.~\cite{vanRitbergen:1997va}):
\begin{eqnarray}
&&
\frac{d_F^{abcd}d_F^{abcd}}{N_A}\ =\ \frac{N_c^4-6N_c^2+18}{96N_c^2}\,,\qquad\quad
\frac{d_F^{abcd}d_A^{abcd}}{N_A}\ =\ \frac{N_c(N_c^2+6)}{48}\,,\label{d44FA}\\[1mm]
&&
T_F\ =\ \frac12\,,\qquad
C_F\ =\ \frac{N_c^2-1}{2N_c}\,,\qquad
C_A\ =\ N_c\,,\qquad
N_A\ =\ N_c^2-1\,.\label{CACF}
\end{eqnarray}
The part of the obtained result, which is proportional to $(n_f)^{i-1}a_s^i$, coincide with the prediction from Ref.~\cite{Gracey:1994nn}.

There are also some predictions~\cite{Kataev:1999bp}\footnote{We thank A.L.~Kataev, who pointed out to us this result.}, coming from
the Pade resummation, which for $\gamma^{(3)}_{{\mathrm{NS}}}(3,n_f=4)$ gives $3480$ or $3450$ and for $\gamma^{(3)}_{{\mathrm{NS}}}(4,n_f=4)$ gives $4211$ or $4207$ depending on the resummations procedure\footnote{Note, that in the four loops a new colour structures (\ref{d44FA}) are appeared, which can disimprove a resummation.}.
Our explicit results for the four active quarks (with $N_c=3$ and $n_f=4$) are given by:
\begin{eqnarray}
\gamma^{4-loop}_{{\mathrm{NS}}}(3,n_f=4) &=& 5.55556\,\als + 50.39095\, \als^2 + 418.17201\, \als^3 + 4322.89048 \, \als^4\,, \\
\gamma^{4-loop}_{{\mathrm{NS}}}(4,n_f=4) &=& 6.97778\, \als + 60.07233\, \als^2 + 502.91174 \,\als^3 + 5066.33924  \,\als^4\,.
\end{eqnarray}
We give also our result for $\gamma^{(3)}_{{\mathrm{NS}}}(3)$ and $\gamma^{(3)}_{{\mathrm{NS}}}(4)$ with the contributions from different colour structures:
\begin{eqnarray}
\gamma^{(3)}_{{\mathrm{NS}}}(3,n_f=4) &=& 4018.37877+364.18329\,  d_{44}^{FA}-59.67157\,  d_{44}^{FF}\\
%
\gamma^{(3)}_{{\mathrm{NS}}}(4,n_f=4) &=& 4968.40347+173.46491\,  d_{44}^{FA}-75.52914\,  d_{44}^{FF}
\end{eqnarray}
where $d_{44}^{FF}$ and $d_{44}^{FA}$ are the contributions coming from the first and second structures in~(\ref{d44FA}) correspondingly.
One can see, that the contribution from $d_{44}^{FF}$ and $d_{44}^{FA}$ become small with compare to the total results with growth of $N$ and closer to the predicted values from Ref.~\cite{Kataev:1999bp}.

In the end we want to give some general results, which can be obtained from our results for $N=2$~\cite{Velizhanin:2011es}, $N=3$~(\ref{ADM3}) and $N=4$~(\ref{ADM4}).
We have found two different colour structures, which have multiplied to some common prefactors with arbitrary $N$:
\begin{enumerate}
  \item[1)] with prefactor equal to the one-loop non-singlet anomalous dimension \cite{Gross:1973ju}:
\begin{eqnarray}
a\, {\CF} \left[2\,S_1+\frac{2}{N(N+1)}-3\right]&\times&
\Bigg(
1
+16\, a^2 {\CF} {\nf} {\TF} {\z3}
-16\, a^2 {\CA} {\nf} {\TF} {\z3}\nonumber\\ &&\hspace*{-10mm}
+\frac{128}{27}\, a^3 {\TFP3} {\nfP3} {\z3}
+\ 32\, a^3 {\CF} {\TFP2} {\nfP2} {\z4}
-32\, a^3 {\CA} {\TFP2} {\nfP2} {\z4}\nonumber\\ &&\hspace*{-10mm}
-160\, a^3 {\CFP2}  {\TF} {\nf} {\z5}
+\frac{80}{3}\, a^3 {\CF} {\CA} {\TF} {\nf} {\z5}
\Bigg).\label{GenI}
\end{eqnarray}
  \item[2)] with prefactor, which can be read form the general expression for three-loop non-singlet anomalous dimension~\cite{Moch:2004pa}:
\begin{eqnarray}
24\,a^3 {\CF}\!
\left[-2\,S_{-2}+\frac{(-1)^N}{N^2(N+1)^2}+\frac{(-1)^N}{N(N+1)}-\frac{5}{4}\right]&\!\times\!&
\Bigg(
4\, {\CFP2} {\z3}
-6\, {\CF} {\CA} {\z3}
+2\, {\CAP2} {\z3}\nonumber\\ &&\hspace*{-60mm}
+8\, a\, {\CFP2} {\TF} {\nf} {\z4}
-22\, a\, {\CFP2} {\CA} {\z4}
+33\, a\, {\CF} {\CAP2} {\z4}
-11\, a\, {\CAP3} {\z4}
\Bigg).\qquad\label{GenII}
\end{eqnarray}
\end{enumerate}
Here we used the nested harmonic sums defined as (see \cite{Vermaseren:1998uu,Blumlein:1998if}):
\begin{eqnarray}
S_a (N)&=&\sum^{N}_{j=1} \frac{(\mbox{sgn}(a))^{j}}{j^{\vert a\vert}}\, , \qquad\qquad
S_{a_1,\ldots,a_n}(N)\ =\ \sum^{N}_{j=1} \frac{(\mbox{sgn}(a_1))^{j}}{j^{\vert a_1\vert}}
\,S_{a_2,\ldots,a_n}(j)\label{vhs}
\end{eqnarray}
and the sum of the absolute values of the indices is called the \emph{transcendentality}:
\begin{equation}
\ell=\vert a_1 \vert +\ldots \vert a_n \vert\,.
\end{equation}

A general results for the anomalous dimension can be obtained also with the method, which was proposed by us for the reconstruction of a general form of anomalous dimension from the few first fixed values. This method based on the two observation:
\begin{enumerate}
  \item[1)] the anomalous dimension of twist-2 operators contains only harmonic sums with a transcendentality not higher than $(2\ell-1)$ at $\ell$-loop order;
  \item[2)] the coefficients in the front of this sums usually rather simple integer numbers.
\end{enumerate}
So, we can write down ansatz from the all possible harmonic sums and find coefficients in the front of these sums from the knowledge of the anomalous dimension with the fixed values. According to the second observation the obtained system of linear equations will be the system of Diophantine equations, which is very interesting from the point of view a number theory. The most simple and the most public available method for the solution of the system of linear Diophantine equations based on the \texttt{LLL}-algorithm~\cite{Lenstra:1982}, which has a realizations in a lot of computer algebra system and a private codes. Firstly, we used this method in our paper~\cite{Velizhanin:2010cm} about a six-loop anomalous dimension of twist-3 operators in $\cN=4$ SYM theory, where we reconstruct the general form of the anomalous dimension from the knowledge of the first 25 even fixed values for the ansatz with 85 harmonic sums. Then we used this method for the reconstruction of a general form of the three-loop anomalous dimension of the transverse twist-2 operators in QCD from the first 15 fixed values~\cite{Velizhanin:2012nm} and the obtained result was checked with 16th moment~\cite{Bagaev:2012bw}. The detailed explanation of the work of this method for our purpose with a simple example can be found in our paper~\cite{Velizhanin:2013vla}.
Recently, we reconstruct with our method a full planar six-loop anomalous dimension of twist-2 operators in $\cN=4$ SYM theory form the first 35 even values~\cite{Marboe:2014sya} (note, that the corresponding ansatz contained about 350 harmonic sums, that is in one order more than the number of calculated fixed values). Not long ago, our method was used for the reconstruction of the three-loop anomalous dimension of the polarized twist-2 operators in QCD from the first 15 odd values~\cite{Vogt:2014pha,Moch:2014sna}.
 So, this method is very powerful, but according to its applicability it demands the usage of a large numbers, that is it will give the more reliable result when the available fixed values would be a large numbers. In this sense the results, presented in this paper, do not contain a large numbers, so we can try to reconstruct only the most simple contribution from the full anomalous dimension, namely the contribution, which is proportional to $\z5$.

There are nine different color structures in our results~(\ref{ADM3}) and (\ref{ADM4}) for $\z5$ contribution:
\begin{eqnarray}
&&
\Bigg\{
a^4 {\CFP4},
a^4 {\CFP3} {\CA},
a^4 {\CFP2} {\CAP2},
a^4 {\CF} {\CAP3},
a^4 {\CFP3}\TF {\nf},
a^4 {\CFP2} {\CA}\TF {\nf},
a^4 {\CF} {\CAP2} \TF{\nf},\nonumber\\[2mm]&&\hspace*{6mm}
a^4 \, \frac{{d_F^{abcd}d_A^{abcd}}}{\NF},
a^4 \TF {\nf} \, \frac{{d_F^{abcd}d_F^{abcd}}}{{\NF}}
\Bigg\}\,.
\end{eqnarray}
The contributions, which are proportional to $a^4 {\CFP3} {\nf}$ and $a^4 {\CFP2} {\CA} {\nf}$ can be found in Eq.~(\ref{GenI}).
The most simple contribution among the rest structures, as we can expect, is proportional to $a^4 {\CF} {\CAP2} \TF{\nf}$.

For the reconstruction we can use also well known fact that the non-singlet anomalous dimension at $N=1$ is equal to zero in all order of perturbative theory, because in this case our operator~(\ref{NSOpN}) is equivalent to a non-singlet vector current and the vanishing of the first moment of the non-singlet anomalous dimension follows from the conservation of the such current.
So, this will give us additional constraints.
Another constraint comes from the results, which are known in $\cN=4$ SYM theory. At present we know a general expression for the anomalous dimension of twist-2 operators up to six-loops for planar part~\cite{Bajnok:2008qj,Lukowski:2009ce,Marboe:2014sya} and at least $\z5$ contribution for non-planar part at four-loop order~\cite{Velizhanin:2009gv,Velizhanin:2010ey,Velizhanin:2014zla}. At four loops the general expression for the $\z5$ contribution can be written as~\cite{Velizhanin:2009gv,Velizhanin:2010ey,Velizhanin:2014zla}: 
\begin{eqnarray}
\gamma^{{\mathrm {4-loop}},\,\z5}_{\cN=4 {\mathrm{SYM}}}(j)&\ =&\ -640\; S_1^2(j-2)\left(1+\frac{12}{N_c^2}\right)\,\z5\,. \label{resnpuad}
\end{eqnarray}
This result should be obtained from the corresponding result in QCD, which we are looking for, applying to its the maximal transcendentality principle, as was done by us in Ref.~\cite{Kotikov:2004er} at three loops. To move from QCD to $\cN=4$ SYM theory we should make arguments of all harmonic sums the same (in general the QCD results contain the harmonic sums with a shifted arguments, see~\cite{Moch:2004pa}), that is change $(N\pm 1)$ to $N$ and make the following substitution for the colour factors: $C_F \to C_A$, $T_F n_f\to C_A$ and $d_F\to d_A$. Then we should drop out all harmonic sums, which do not respect to the maximal transcendentality principle, that is all harmonic sums with transcendentality less than $(2\ell-1)$ at $\ell$-loop order. Following the maximal transcendentality principle we can expect the next harmonic sums in the expression for the $\z5$ contribution to the four-loop non-singlet anomalous dimension:
\begin{eqnarray}
&&
\bigg\{
S_{1,1}(N)-\frac{1}{2}S_2(N),
S_{1,1}(N+1)-\frac{1}{2}S_2(N+1),
S_{1,1}(N+1)-\frac{1}{2}S_2(N+1),\nonumber\\[0mm]&&\hspace*{20mm}
S_{-2}(N),
S_{-2}(N+1),
S_{-2}(N-1),
S_{1}(N),
S_{1}(N+1),
S_{1}(N-1),
1
\bigg\}\,,\label{Basis}
\end{eqnarray}
where the combination of the harmonic sums in the first three terms and the absence of $S_2$ sums are due to reciprocity~\cite{Dokshitzer:2005bf,Dokshitzer:2006nm}, which, as we believe, should works for QCD in this case.
Applying above described procedure to the result, which will some combinations of the harmonic sums form the basis~(\ref{Basis}) we should obtain the result from Eq.~(\ref{resnpuad}).
Note, that if we make the substitution for the colour factors $C_F \to C_A$ in our results for $N=2,\,3,\,4$ we obtain:
\begin{eqnarray}
\gamma^{(3)}_{{\mathrm{NS}},\,\z5}(2)& \overset{\CF\to\CA}{=}\ &
\frac{1280}{27}
\left(12\, \frac{{d_F^{abcd}d_A^{abcd}}}{{\NF}}+\CAP4\right)
-\frac{640}{27}
{ {\nf} \left(12\, \frac{{d_F^{abcd}d_F^{abcd}}}{{\NF}}+\CAP3\TF\right)},\label{relatpnp1}\\
\gamma^{(3)}_{{\mathrm{NS}},\,\z5}(3)& \overset{\CF\to\CA}{=}\ &
-\frac{1825 }{27}
\left(12\, \frac{{d_F^{abcd}d_A^{abcd}}}{{\NF}}+\CAP4\right)
-\frac{1000}{27}
{ {\nf} \left(12\, \frac{{d_F^{abcd}d_F^{abcd}}}{{\NF}}+\CAP3\TF\right)},\label{relatpnp2}\\
\gamma^{(3)}_{{\mathrm{NS}},\,\z5}(4)& \overset{\CF\to\CA}{=}\ &
-\frac{19717}{135}
\left(12\, \frac{{d_F^{abcd}d_A^{abcd}}}{{\NF}}+\CAP4\right)
-\frac{6728}{135}
{ {\nf} \left(12\, \frac{{d_F^{abcd}d_F^{abcd}}}{{\NF}}+\CAP3\TF\right)}\,.\qquad\label{relatpnp3}
\end{eqnarray}
So, indeed, planar and non-planar contributions are related through the same equations as we have in $\cN=4$ SYM theory, even for $n_f$ part.

We start our attempts of reconstruction with the colour structures, which are proportional to the $n_f$. For such contributions we can make some interesting observation from the known three-loop results~\cite{Moch:2004pa}: the transcendentality of such contributions is less than the maximal transcendentality on the power of $n_f$. In our case we can expect, that all contributions, which are proportional to $n_f$ will have transcendentality $1$. For $a^4 {\CFP3} {\nf}$ and $a^4 {\CFP2} {\CA} {\nf}$ this is correct~(\ref{GenI}). So, for $a^4 {\CF} {\CAP2} \TF{\nf}$ and $a^4 {\nf}\, {{d_F^{abcd}d_F^{abcd}}}\!/\!{{\NF}}$ we will write the following basis (compare with Eq.~(\ref{GenI})):
\begin{eqnarray}
&&
\bigg\{
S_{1}(N)\,,\
\frac{1}{N(N+1)}\,,\
\frac{(-1)^N}{N(N+1)}\,,\
1
\bigg\}\,.\label{Basisd44nf}
\end{eqnarray}
As we have four results for $N=1,\,2,\,3,\,4$ we can exactly find all coefficients in the ansatz from the basis~(\ref{Basisd44nf}) and the corresponding contributions look like:
\begin{eqnarray}&&
\gamma^{(3)}_{{\mathrm{NS}},\,\z5}\Big|_{{{d_Fd_F}}}=
\frac{1024}{3}\,\TF\, {\nf} \, \frac{{d_F^{abcd}d_F^{abcd}}}{{\NF}}
\bigg(
-7 S_1(N)
-\frac{5}{2 N (N+1)}
-\frac{(-1)^N}{N (N+1)}
+\frac{31}{4}
\bigg),\qquad\label{Res1d44nf}\\&&
\gamma^{(3)}_{{\mathrm{NS}},\,\z5}\Big|_{\CF\CAP2\nf}=
\frac{128}{9} \CF \CAP2 \TF \nf
\bigg(
17 S_1(N)
-\frac{10}{N (N+1)}
-\frac{(-1)^N}{4 N (N+1)}
-\frac{97}{8}
\bigg)\,.\qquad\label{Res1CFnf}
\end{eqnarray}

For the reconstruction of the non-$\nf$ contributions we tried the following minimal basis:
\begin{eqnarray}
&&
\bigg\{
S_{1}(N)^2\,,\
S_{1}(N)\,,\
\frac{1}{N^2(N+1)^2}\,,\
\frac{(-1)^N}{N^2(N+1)^2}\,,\
\frac{1}{N(N+1)}\,,\
\frac{(-1)^N}{N(N+1)}\,,\
1
\bigg\}\,.\label{Basisd44}
\end{eqnarray}
From the corresponding ansatz and with the help of \texttt{LLL}-algorithm we have found the following result for the non-planar contribution:
\begin{eqnarray}
\gamma^{(3)}_{{\mathrm{NS}},\,\z5}\Big|_{{{d_Fd_A}}}&=&
640 \, \frac{{d_F^{abcd}d_A^{abcd}}}{{\NF}}
\bigg(
-4 S_1^2(N)
+2 S_1(N)
-\frac{3}{N^2 (N+1)^2}
+\frac{10 (-1)^N}{N^2 (N+1)^2}\nonumber\\[0mm]&&\hspace*{30mm}
-\frac{3}{N (N+1)}
+\frac{2 (-1)^N}{N (N+1)}
+\frac{31}{4}
\bigg)\,,\label{Res2d44}
\end{eqnarray}
which coincides with $\cN=4$ SYM theory limit~(\ref{resnpuad}) take into account a difference in normalization and the following substitution $d_F^{abcd}d_A^{abcd}/\NF\to 3/2 N_c^2$.
For other contributions the most simple expressions, obtained with the help of \texttt{LLL}-algorithm from the minimal basis~(\ref{Basisd44}) are the following:
\begin{eqnarray}
\gamma^{(3)}_{{\mathrm{NS}},\,\z5}\Big|_{\CFP4}&=&
320\, \CFP4
\bigg(
+4 S_1^2(N)
-14 S_1(N)
+\frac{14 (-1)^N}{N^2 (N+1)^2}\nonumber\\[0mm]&&\hspace*{30mm}
+\frac{10}{N^2 (N+1)^2}
-\frac{(-1)^N}{N (N+1)}
+\frac{1}{N (N+1)}
+10
\bigg)\,,\label{Res2cf4}\\
\gamma^{(3)}_{{\mathrm{NS}},\,\z5}\Big|_{\CFP3\CA}&=&
320\, \CFP3\CA
\bigg(
-2 S_1^2(N)
+6 S_1(N)
+\frac{5 (-1)^N}{N^2 (N+1)^2}
+\frac{4}{N^2 (N+1)^2}\nonumber\\[0mm]&&\hspace*{30mm}
-\frac{5 (-1)^N}{N (N+1)}
-\frac{11}{N (N+1)}
-\frac{3}{4}
\bigg)\,,\label{Res2cf3ca}\\
\gamma^{(3)}_{{\mathrm{NS}},\,\z5}\Big|_{\CFP2\CAP2}&=&
160\, \CFP2\CAP2
\bigg(
-S_1^2(N)
-S_1(N)
+\frac{5 (-1)^N}{2 N^2 (N+1)^2}\nonumber\\[0mm]&&\hspace*{30mm}
+\frac{17 (-1)^N}{2 N (N+1)}
+\frac{9}{2 N (N+1)}
+\frac{37}{8}
\bigg)\,,\label{Res2cf2ca2}\\
\gamma^{(3)}_{{\mathrm{NS}},\,\z5}\Big|_{\CF\CAP3}&=&
\frac{160}{3}\, \CF\CAP3
\bigg(
+4 S_1^2(N)
-4 S_1(N)
-\frac{10 (-1)^N}{N^2 (N+1)^2}
+\frac{3}{N^2 (N+1)^2}\nonumber\\[0mm]&&\hspace*{30mm}
-\frac{7 (-1)^N}{N (N+1)}
+\frac{10}{N (N+1)}
-\frac{47}{4}
\bigg)\,.\label{Res2cfca3}
\end{eqnarray}
However Eqs.~(\ref{Res2cf4})-(\ref{Res2cfca3}) do not satisfy the relation from Eqs.~(\ref{relatpnp1})-(\ref{relatpnp3}). In principal, we can extended basis~(\ref{Basisd44}), but our numbers from Eqs.~(\ref{ADM3}) and~(\ref{ADM4}) are rather small and we can not improve our general results. So, we need to know the higher moments.

In general we can conclude, that the structure of the anomalous dimension in the four loops is much complicated with compare to what we can expect from the analyze of the three-loop results.
Note, that something similar occur in $\cN=4$ SYM theory starting from the four loops.
Up to three loops the anomalous dimension of the twist-2 operators can be obtained with the help of integrability from the asymptotic Bethe-ansatz (ABA)~\cite{Staudacher:2004tk}, while starting from the four loops a new types of diagrams are appeared, which are not accounted with ABA~\cite{Kotikov:2007cy}. We are going to continue our calculations to obtain a general result, at least for the $\z5$ contribution.

 \subsection*{Acknowledgments}
We would like to thank A.L.~Kataev, L.N.~Lipatov and A.I.~Onishchenko for useful discussions. We thank S. Moch for pointing out to us misprints in Eqs.~(\ref{relatpnp1})-(\ref{relatpnp3}).
The research of V.N.~Velizhanin is supported by a Marie Curie International Incoming Fellowship within the 7th European Community Framework Programme, grant number PIIF-GA-2012-331484, by DFG SFB 647 ``Raum -- Zeit -- Materie. Analytische und Geometrische Strukturen'' and by RFBR grants 12-02-00412-a, 13-02-01246-a.

\end{document}